\documentclass{elsart}
\usepackage{amssymb}

\renewcommand{\d}{{\mathrm d}}
\renewcommand{\bar}[1]{\overline{#1}}

\usepackage{indentfirst}
\usepackage{psfig,color}
\usepackage{epsfig}
\usepackage{epsf}
\usepackage{graphicx}
\journal{Physics Letters B}

\begin{document}

\begin{frontmatter}
\title{Sigma Meson Cloud \\ And Proton'S Light Flavor Sea Quarks}

%note to the publisher: please keep the institutions as here and dont make change
\author[pku]{Feng Huang},
\author[pku]{Rong-Guang Xu},
\author[ccastpku]{Bo-Qiang Ma\corauthref{cor}}
\corauth[cor]{Corresponding author.} \ead{mabq@phy.pku.edu.cn}
\address[pku]{Department of Physics, Peking University, Beijing 100871, China}
\address[ccastpku]{CCAST (World Laboratory), P.O.~Box 8730, Beijing
100080, China\\
Department of Physics, Peking University, Beijing 100871, China}

\begin{abstract}
We take into account the sigma meson cloud effect in the meson cloud model to calculate
the distributions of light flavor sea quarks in the proton. Our
calculation gives a better description of the data for
$\bar{d}(x)/\bar{u}(x)$. We also provide a picture that the
probability of finding a physical proton in a Fock state
$\left|N\omega\right> $ is reasonable small with a smaller cutoff
$\Lambda_{\omega}$.
\end{abstract}

\end{frontmatter}

\par%%% Introduction------------------------------------------------------
By now, it has been experimentally established that the light-flavor sea quarks
$\bar{d}(x)$ and $\bar{u}(x)$ in the proton differ substantially~\cite{NMC,KA,AB,E866}.
The violation of the Gottfried sum rule,
first found by NMC~\cite{NMC}, indicated that
$D\equiv\int_0^1[\bar{d}(x)-\bar{u}(x)]\,\d x =0.148\pm0.039$,
and similar result for $\bar{d}(x)-\bar{u}(x)$ was obtained by
HERMES \cite{KA}. In Drell-Yan experiments, $\bar{d}/\bar{u}$ was
determined to be more than 2 at $x=0.18$ by NA51~\cite{AB} and the
Bjorken-$x$ dependence~(moment fraction dependence) of the ratio
has been measured by E866/NuSea~\cite{E866}.

While no known symmetry requires $\bar{d}(x)$ equal to
$\bar{u}(x)$, the large asymmetry was out of naive expectations.
The sea of light flavor quark-antiquark pairs produced
perturbatively from gluon splitting is flavor symmetric. Thus the
large asymmetry requires a nonperturbative origin. It was first
predicted by Thomas \cite{first}, and later suggested by Henley and
Miller~\cite{first2}, and Kumano and Londergan~\cite{SK}, that
including the effects of virtual mesons~(recognized as meson cloud
model) can naturally account for the %observed 
asymmetry. In
previous articles applying meson cloud model to explain the light
flavor sea asymmetry in the nucleon, a good description of the
data was obtained for the distribution functions \cite{JS}. The
dominant role is played by the pion, which provides that
$\bar{d}(x)/\bar{u}(x)$ either increases monotonically with $x$ or
turns back towards unity too slowly \cite{JS,MA}. Many
explanations have been applied to attack this problem
\cite{AS,WK,WM,NN}, such as effects of $\Delta$ \cite{AS,WK,WM},
the influence of the Pauli exclusion principle \cite{WM},
adjustment of parameters \cite{NN}, but none of these explanations
provides a satisfactory description of the ratio
$\bar{d}(x)/\bar{u}(x)$. However, Alberg, Henley and
Miller~\cite{omega} found that the inclusion of the effect of the
isoscalar vector meson $\omega$, with a coupling constant
$g_{\omega}^2/4\pi\approx8.1$, allows a good
description of the data. As mentioned in that article, adding the
$\sigma$ effect will likely improve the description of the data.
Here, we follow this suggestion and take into account the sigma
meson effect in the meson cloud model.

The existence of the $\sigma$ meson has been obscure for many years,
while many theorists and experimentalists were searching for
$\sigma$ which may play an important role in nuclear physics,
because the $\sigma$ meson can provide reasonable middle-range
nuclear force \cite{meson}. Sigma meson was first introduced as
the chiral partner of pion in the linear representation of chiral
symmetry, which is an important ingredient in modern hadron
physics. It also plays an important role in spontaneous chiral
symmetry breaking to understand the present spectroscopy of
hadrons. However, it was not well established experimentally
mainly due to the negative results of extensive analysis of the
$\pi$-$\pi$ scattering phase shift. While recent reanalysis of the
$\pi$-$\pi$ scattering phase shift \cite{OP} strongly
suggested a pole of the $\sigma$-meson. Though we do not have a
confirmed conclusion that the sigma meson exists as a real
physical particle, it is sufficient for considering its effect to
calculate $\bar{d}(x)-\bar{u}(x)$ and $\bar{d}(x)/\bar{u}(x)$ in
the proton in the meson cloud model.

In the meson cloud model \cite{SK,JS,MA}, the nucleon can be
viewed as a bare nucleon~(core) plus a series of baryon-meson Fock states
which result from the fluctuation of nucleon to baryon plus meson
$N\rightarrow BM$~(a bare baryon surrounded by a meson cloud). The
model assumes that the life-time of a virtual baryon-meson Fock
state is much longer than the interaction time in deep
inelastic scattering~(DIS) or Drell-Yan processes, thus the quark
and anti-quark in the virtual baryon-meson Fock states can
contribute to the parton distribution of the nucleon. These
non-perturbation contribution can be calculated in a convolution
between the fluctuation function, which describes the microscopic
process $N\rightarrow BM$, and the quark~(anti-quark) distribution
of hadrons in the Fock states $\left|BM\right> $. Here we
provide the usual formula. First the physical proton wave function
is composed of the following Fock states
\begin{eqnarray}%(1)
\left|P\right>=\sqrt{Z}\left|P\right>_{\tiny{\textrm{bare}}}+\sum_{BM}\int
\d y \d ^{2}\textbf{k}_{\perp}\Psi_{BM}(y,\textbf{k}_{\bot}^{2})
\left|B(y,\emph{\textbf{k}}_{\bot}),M(1-y,-\emph{\textbf{k}}_{\bot})\right>\nonumber.
\end{eqnarray}
Here $Z$ is the wave function renormalization constant,
$\Psi_{BM}(y,\textbf{k}_{\bot}^{2})$ is the probability amplitude for
finding a physical nucleon in a state consisting of a baryon $B$
with longitudinal moment fraction $y$ and transverse momentum
$\emph{\textbf{k}}_{\bot}$, and a meson $M$ with longitudinal moment
fraction $1-y$ and transverse momentum $-\emph{\textbf{k}}_{\bot}$.
Therefore, the quark distribution functions $q(x)$ in the proton
are given by
\begin{eqnarray}%(2)
q(x)=Z q_{\tiny{\textrm{bare}}}(x)+\delta q(x)\;,
\end{eqnarray}
with
\begin{eqnarray}%(3)
\delta q(x)=\sum_{MB}[\int_x^1\frac{\d
y}{y}f_{MB}(y)q_{M}(\frac{x}{y})+\int_x^1\frac{\d
y}{y}f_{BM}(y)q_{B}(\frac{x}{y})]\;,
\end{eqnarray}
and
\begin{eqnarray}%(4)
f_{MB}(y) = f_{BM}(1-y)\;.\label{sym}
\end{eqnarray}
As shown in \cite{WK}, the wave function renormalization constant reads
\begin{eqnarray}%(5)
Z=[1+\sum_{BM}\int_0^1f_{BM}(y)\d y]^{-1}\simeq1-\sum_{BM}\int_0^1f_{BM}(y)\d y\;.
\end{eqnarray}
The number of each type of meson, $n_{M}$, is obtained by
integrating $f_{MB}(y)$ over $y$. Then we have $Z=1-\sum_{M}n_M$.
The splitting function is
\begin{eqnarray}%(6)
f_{BM}(y)=\int_0^\infty|\Psi_{BM}(y,\textbf{k}_{\bot}^{2})|^{2}\d^{2}\textbf{k}_{\bot}\;.
\end{eqnarray}
We use time-ordered perturbation theory~(TOPT) in the infinite
momentum frame~(IMF) to calculate this function, which was given
by~\cite{JS} as
\begin{eqnarray}%(7)
f_{BM}(y)=\frac{1}{4\pi^{2}} \frac{m_{N}m_{B}}{y(1-y)}
\frac{|G_{M}(y,\textbf{k}_{\bot}^2)|^{2}|V_{\mathrm{IMF}}|^{2}}{[m_N^2-M_{BM}^2(y,\textbf{k}_{\bot}^2)]^2}\;,
\end{eqnarray}
where
\begin{eqnarray}%(8)
M_{BM}^2(y,\textbf{k}_{\bot}^2)=\frac{m_B^2+\textbf{k}_{\bot}^2}{y}+\frac{m_M^2+\textbf{k}_{\bot}^2}{1-y}
\end{eqnarray}
is the invariant mass of the final state, and an exponential form
for the cutoff is \cite{WK,meson}
\begin{eqnarray}%(9)
G_{M}(y,\textbf{k}_{\bot}^2)=\exp(\frac{m_N^2-M_{BM}^2(y,\textbf{k}_{\bot}^2)}{2\Lambda_M^2})\;,
\end{eqnarray}
which is required to respect Eq.(\ref{sym}). Here $\Lambda_M$ is a
cutoff parameter for each meson.

The vertex function $V_{\mathrm{IMF}}(y,\textbf{k}_{\bot}^2)$ depends on
the effective interaction Lagrangian that describes the
fluctuation process $N\rightarrow BM$. Here we use the meson
exchange model for hadron production~\cite{meson} and the linear
sigma model~\cite{book} to perform the calculation.

In a full calculation, we should include all kinds of mesons and
baryons. While the probability of baryon-meson fluctuation should
decrease with the invariant mass of the baryon-meson Fock state
increasing, we can neglect the effects of Fock states with higher invariant mass. We
just include specifically $\pi$, $\sigma$ and $\omega$ mesons with
the nucleon here. It has been discussed that the effect of the $\rho$
meson and the intermediate $\Delta$ can also contribute a lot
each. But the $\rho$ meson increases $\bar{d}(x)-\bar{u}(x)$,
whereas the intermediate $\Delta$ decreases it, so these effects
tend to cancel each other. And we also omit the effect of $\eta$ for
its small coupling constant. Thus it is safe to focus on the
effects of $\pi$, $\sigma$ and $\omega$ meson clouds here.

First, we focus on the sigma meson. From the Bonn meson-exchange
model for the nucleon-nucleon interaction \cite{meson}, the
effective interaction Lagrangian for the scalar meson $\sigma$ is
$\emph{L}=g\bar{\psi}\sigma\psi$. Thus we can get the vertex
function for $\sigma$ to obtain the fluctuation function as
\begin{eqnarray}%(10)
f_{N\sigma}(y)=\frac{g_{\sigma}^2}{8\pi^2} \frac{1}{y^2(1-y)}
\int_0^\infty \d \textbf{k}_{\bot}^2 |G_{\sigma}(y,\textbf{k}_{\bot}^2)|^2
\frac{m_N^2(1-y)^2+\textbf{k}_{\bot}^2}{[m_N^2-M_{N\sigma}^2(y,\textbf{k}_{\bot}^2)]^2}\;.
\end{eqnarray}%(10)
Now, we should know the coupling constant $g_\sigma$, the mass $m_\sigma$
and the cutoff $\Lambda_\sigma$.

As mentioned above, sigma meson is introduced to the hadron
physics as the chiral partner from the view of chiral symmetry. In
the $SU(2)$ linear $\sigma$ model \cite{book}, the effective
Lagrangian is
\begin{eqnarray}%(11)
\emph{L}=g\bar{\psi}(\sigma+\gamma_5\mathbf{\tau}\cdot\mathbf{\pi})\psi\;.
\end{eqnarray}
Even though we do not have direct information about $g_\sigma$,
$g_\sigma = g_\pi$ is imposed by the linear $\sigma$ model.
According to all kinds of calculations to fit different
experimental data \cite{meson,RJ,range}, the coupling constant
$g_\sigma^2/4\pi=g_\pi^2/4\pi=13.6$ is taken in our numerical
computations. In 2002 Particle Data Group~(PDG), $f_0(600)$ or
$\sigma$ appears below $1~\textrm{GeV}$ mass region. Recently it
was found that $m_\sigma=585\pm20~\textrm{MeV}$ by a reanalysis of
$\pi\pi$ scattering phase shift \cite{OP}. Similarly, by analysing
$\pi\pi$ production processes \cite{OP}, we obtain
$m_\sigma=580\pm30~\textrm{MeV}$ for \emph{pp} central collision.
Thus, we will set the mass as $600~\textrm{MeV}$ in the following
calculations. It is generally believed that the cutoff
$\Lambda_\sigma$  value is in a range around $1~\textrm{GeV}$ as
a phenomenological parameter. Here we will examine the effect of
varying $\Lambda_\sigma$  value in the range
$1.0<\Lambda_\sigma<1.3~\textrm{GeV}$.

\begin{figure}%(1)
\includegraphics{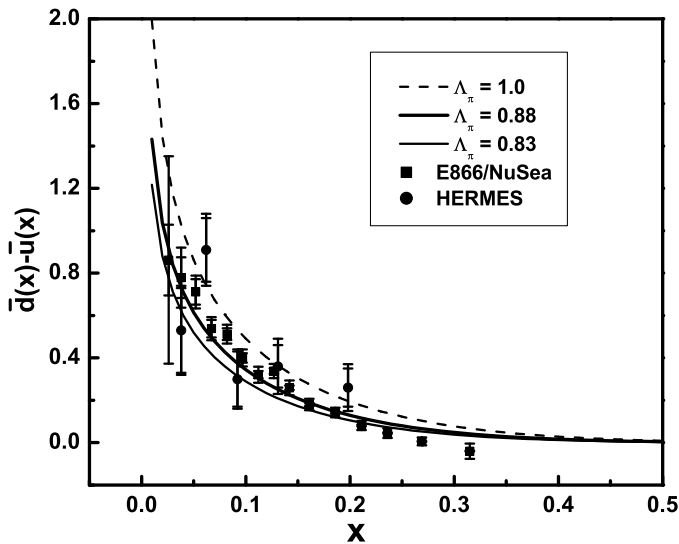}
\caption{\label{fig:d-u} \small  Comparison of our meson cloud model with
data \cite{KA,E866} for $\bar{d}(x) - \bar{u}(x)$. The thick
solid curve is for $\Lambda_\pi = 0.88~\textrm{GeV}$, which gives a
best description here.}
\end{figure}

We have made some detailed discussions about the $\sigma$ meson
cloud above. The $\pi$ meson cloud effect has been discussed
enough in previous articles \cite{first2,JS,omega,HN}, here we
just set the parameters as $m_\pi=139~\textrm{MeV}$ and
$\Lambda_\pi=(0.88\pm0.05)~\textrm{GeV}$, which are chosen to
reproduce the integrated asymmetry
$D\equiv\int_0^1[\bar{d}(x)-\bar{u}(x)]\,dx$ \cite{NMC,KA} as
shown in Fig.~\ref{fig:d-u}.

Moreover, we should discuss the functions of $q_M(x)$ and
$q_B(x)$. Those for the nucleon and pion are measured, but the
quark distribution functions of the  $\sigma$ meson and $\omega$
meson are unknown. It has been traditional \cite{same} to assume
that the structure function of the $\rho$ and $\pi$ are the same.
It is suggested by the bag model \cite{omega} that the structure
functions of the $\omega$, $\rho$, and $\pi$ are the same.  Due to
that the $\sigma$ meson operator is adopted as
$\frac{1}{\sqrt{2}}(\bar{u}u+\bar{d}d)$, there is some reason to
assume that the quark distribution functions of the $\sigma$ meson
is equal to that of the $\pi$. Thus we use \cite{dis}
\begin{eqnarray}%(13)
xq_{\tiny{\textrm{V}}}(x)&=&0.99x^{0.61}(1-x)^{1.02}\nonumber,\\
xq_{\tiny{\textrm{sea}}}(x)&=&0.2(1-x)^{5.0}
\end{eqnarray}%(14)
for the valence and sea quark distribution functions of all mesons
considered here.

\begin{figure}%(2)
\includegraphics{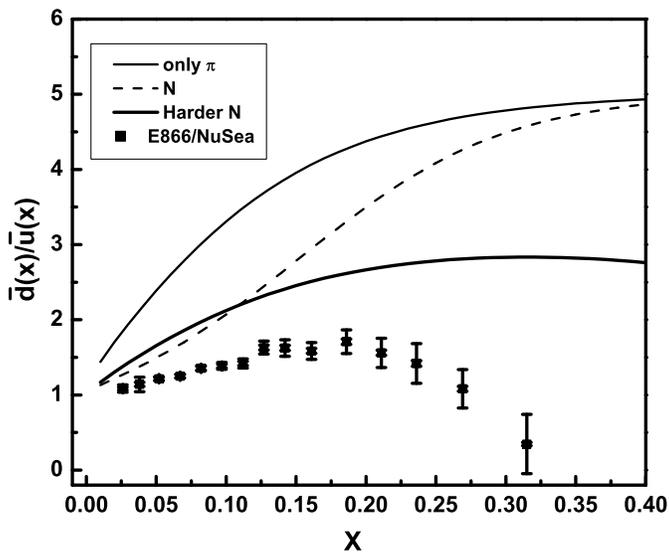}
\caption{\label{fig:HN} \small Comparison of a harder bare nucleon sea quarks~(thick
solid line) with a traditional bare nucleon sea quarks~(dashed
line). The thin solid line is only the $\pi$ contribution to the ratio
$\bar{d}(x)/ \bar{u}(x)$.}
\end{figure}

The bare nucleon sea is traditionally parametrized as \cite{N}
\begin{eqnarray}%(15)
x\bar{q}_{\tiny{\textrm{bare}}}(x)&=&0.11(1-x)^{15.8}\nonumber,\\
\bar{q}_{\tiny{\textrm{bare}}}=u_{\tiny{\textrm{sea}}}&=&\bar{u}_{\tiny{\textrm{sea}}}=d_{sea}=\bar{d}_{\tiny{\textrm{sea}}}\;,
\end{eqnarray}%(16)
which is recognized as Holtmann's parametrization of the bare
nucleon symmetric sea. While Alberg and Henley~\cite{HN} used a
harder distribution for the bare sea quarks of the form found in
the determination of the gluon distribution~\cite{gluon}
\begin{eqnarray}%(17)
x\bar{q}_{\tiny{\textrm{bare}}}(x)=0.0124x^{-0.36}(1-x)^{3.8}\;.
\end{eqnarray}%(17)
Their calculations indicate that the harder distribution for the
bare sea quarks gives a better description of
$\bar{d}(x)/\bar{u}(x)$ as shown in Fig.~\ref{fig:HN}. In the
following calculations, we will use only the harder bare nucleon
sea. Also we can see that the change is not enough to explain the
data, which guarantees the necessary to consider other flavor
symmetric contribution. The inclusion of isoscalar meson $\omega$ with a
reasonable coupling constant produces a similar improvement in
agreement between theory and experiment \cite{omega}, so does the
inclusion of isoscalar meson $\sigma$.

\begin{figure}%(3)
\includegraphics{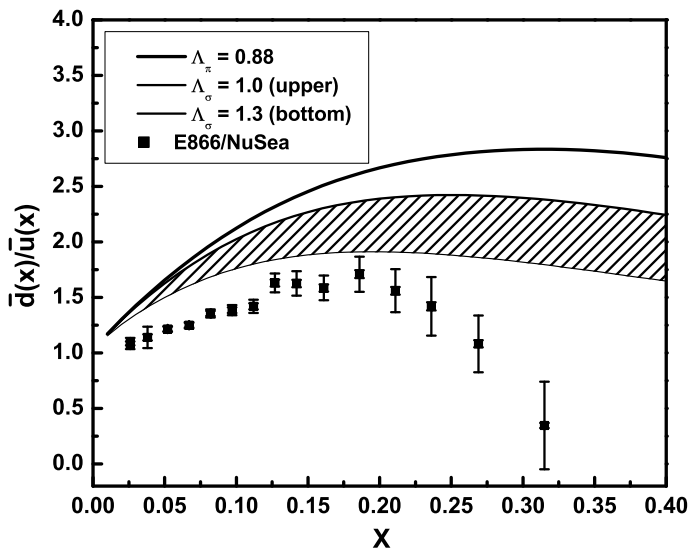}
\caption{\label{fig:ratio2} \small Comparison of our meson cloud model
with data \cite{E866} for $\bar{d}(x)/\bar{u}(x)$.  The thick
solid curve shows our result only considering the $\pi$ meson
contribution. The shadowed area shows the effect of adding
$\sigma$ meson and varying $\Lambda_\sigma$  value in the range
$1.0<\Lambda_\sigma<1.3 \textrm{GeV}$.}
\end{figure}

\begin{figure}[ht]
\begin{minipage}[t]{4.0cm}
\includegraphics{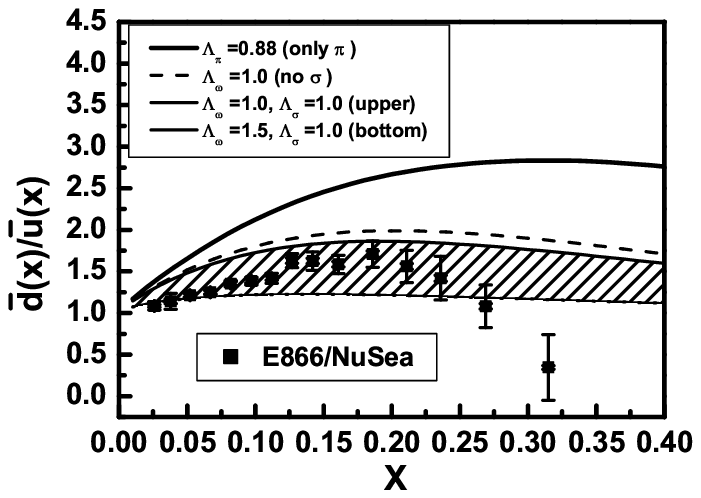}
\end{minipage}
\hfill
\begin{minipage}[t]{7.0cm}
\includegraphics{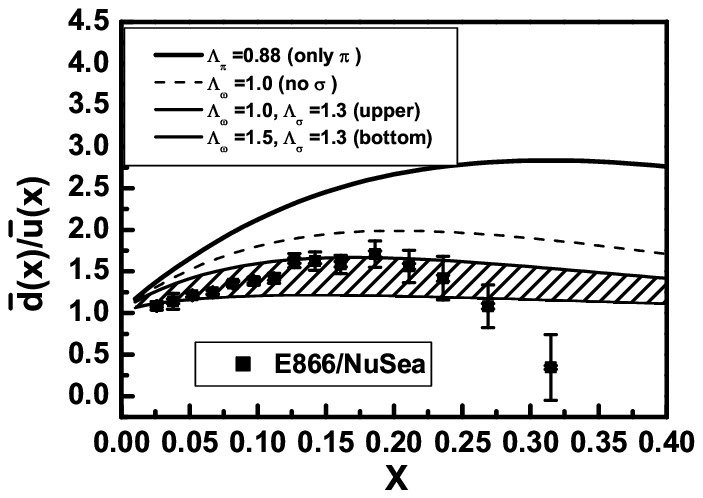}
\end{minipage}
\caption{\label{fig:600} \small  Comparison of our meson cloud model with
data \cite{E866} for $\bar{d}(x)/\bar{u}(x)$. The thick solid
curve shows our result if contributions from the $\sigma$ and
$\omega$ clouds are omitted. The dashed curve shows adding $\omega$
cloud contribution with $\Lambda_\omega=1.0~\textrm{GeV}$. The
shadowed area shows the contributions of three mesons, and the effect
of varying $\Lambda_\omega$  value in the range
$1.0<\Lambda_\omega<1.5 \textrm{GeV}$. }
\end{figure}

%\section{results}
The results of our calculations for light flavor sea quarks are
shown in Fig.~\ref{fig:ratio2} and Fig.~\ref{fig:600}. In
Fig.~\ref{fig:ratio2}, we examine the effect of varying
$\Lambda_\sigma$ value in the range
$1.0<\Lambda_\sigma<1.3~\textrm{GeV}$. The thin solid curve on the
upper of the shadowed area is the result of setting
$\Lambda_\sigma=1.0~\textrm{GeV}$, while the bottom curve  
corresponds to $\Lambda_\sigma=1.3~\textrm{GeV}$. It is clear
that adding sigma meson effect here gives really better
description of the data. Our calculations illustrate that the larger value of
$\Lambda_{\sigma}$ tends to give small values of the ratio
$\bar{d}/\bar{u}$ and that decreasing $\Lambda_{\sigma}$ causes the
maximum value of $\bar{d}/\bar{u}$ to be larger and to appear at
a higher value of $x$. %We also find that the smaller value of
%$m_\sigma$ also tends to give small value of the ratio
%$\bar{d}(x)/\bar{u}(x)$, but the alteration is relatively small
%in the examined range, so we do not show it here.
%\newpage

In Fig.~\ref{fig:600}, we illustrate the effects from 
three mesons with
different cutoff values. As calculated in \cite{omega}, we set
$g_{\omega}^2/4\pi$ as $8.1$, which is from fitting to dispersion
relation descriptions of forward nucleon-nucleon scattering
\cite{meson,range,coupling}. The shadowed area shows the results
of varying $\Lambda_\omega$ value in the range
$1.0<\Lambda_\omega<1.5~\textrm{GeV}$ with  
$\Lambda_\sigma$ at $1.0~\textrm{GeV}$ on the left and
$1.3~\textrm{GeV}$ on the right. The dashed curves omit the
$\sigma$ contribution and only consider the value of
$\Lambda_\omega$ as $1.0~\textrm{GeV}$. When the value of
$\Lambda_\omega$ is $1.5~\textrm{GeV}$, the ratio
$\bar{d}(x)/\bar{u}(x)$ hardly changes with adding $\sigma$ meson
cloud, which is consistent with \cite{omega}, in which 
$\Lambda_\omega$ is set as $1.5~\textrm{GeV}$ to obtain a good
description of data by considering only $\pi$ and $\omega$
effects.
%we set $\Lambda_\sigma$ at $1.0~ \textrm{GeV}$ in the left
%picture, $1.3~\textrm{GeV}$ in the right. The thin solid curve on
%the upper of the shadowed area is the result of setting
%$\Lambda_\omega=1.0~\textrm{GeV}$, while the bottom curve is
%The dashed line is only included $\pi$ and $\omega$ mesons
%contribution. We didn't show
%We find that the better
%description for the data favors the small value of
%$1.0<\Lambda_\omega<1.5~\textrm{GeV}$, which has some modification
%from \cite{omega} as $1.5~\textrm{GeV}$.
In order to have a deep understanding of this, we present the
numbers of the related mesons $n_M$ in the proton with the
different cutoff values in the two works, which are shown in
Table~1. The large value of cutoff leads to a large value of meson
number in the proton. Apparently, the too large $n_\omega$ resulted
from the large value of $\Lambda_\omega=1.5~\textrm{GeV}$ should
not be the real picture of the proton. For this reason,
$\Lambda_\omega$ favors the small value of the examined range.
Adding $\sigma$ cloud effect modifies the parameters describing
the omega-nucleon interaction. And such a revision provides a
picture that the probability of finding a physical proton in a
Fock state $\left|N\omega\right> $ is reasonably small. We also
realize that the same rule holds true for the $\sigma$-meson from
the comparison of the two different values of $\Lambda_\sigma$.
\begin{flushleft}
\begin{tabular}[t]{|c|c|c|c|c|c|c|}
\multicolumn{7}{l}{Table 1. Parameters and meson numbers in the
proton}
\\ \hline \ & $\Lambda_{\pi} $& $\Lambda_{\sigma}$ &
$\Lambda_{\omega}$ & $n_{\pi}$ & $n_{\sigma}$ & $n_{\omega}$ \\
\hline \small{Our work} & $0.88$ & $1.0\sim1.3$ & $1.0\sim1.5$ &
 $0.175$ & $0.023\sim 0.078$ & $0.063\sim 0.671$ \\ \hline
\small{Alberg-Henley's work} & $0.83$ & no sigma & $1.5$ &
$0.150$ & no sigma
 &0.671 \\ \hline
\end{tabular}
\end{flushleft}

%\newpage
%\section{summary}
In summary, the inclusion of the $\sigma$ meson cloud effect brings a better 
description for $\bar{d}(x)/\bar{u}(x)$ in the proton and
also provides a picture of a reasonable small $n_\omega$ in the
proton. Thus we conclude that the inclusion of both isoscalar meson sigma and omega cloud effects
can give an improved description of the experimental data for $\bar{d}(x)/\bar{u}(x)$ in the meson cloud model.

%\section{acknowledgement}
This work is partially supported by National Natural Science
Foundation of China under Grant Numbers 10025523 and 90103007.

%\parbox{10cm}

\end{document}